\documentclass[12pt, preprint]{aastex}

\begin{document}

\shorttitle{Synchronous Optical and Radio Polarization in OJ287}
\shortauthors{D'Arcangelo et al.}

\title{Synchronous Optical and Radio Polarization Variability in the Blazar OJ287} 

\author{Francesca D. D'Arcangelo\altaffilmark{1}, Alan P. Marscher\altaffilmark{1},
Svetlana G. Jorstad\altaffilmark{1,2}, Paul S. Smith\altaffilmark{3}, Valeri M. Larionov\altaffilmark{2,4}, 
Vladimir A. Hagen-Thorn\altaffilmark{2,4}, G. Grant Williams\altaffilmark{5}, 
Walter K. Gear\altaffilmark{6}, Dan P. Clemens\altaffilmark{1}, Domenic Sarcia\altaffilmark{1}, Alex Grabau\altaffilmark{1}, Eric V. Tollestrup\altaffilmark{1}, Marc W. Buie\altaffilmark{7}, Brian Taylor\altaffilmark{7}, and Edward Dunham\altaffilmark{7}}

\altaffiltext{1}{Institute for Astrophysical Research, Boston University, Boston, MA 02215; fdarcang@bu.edu, marscher @bu.edu, jorstad@bu.edu, clemens@bu.edu, dsscryo@gmail.com, grabau@gmail.com, tolles@ifa.hawaii.edu.}
\altaffiltext{2}{Sobolev Astronomical Institute, St. Petersburg State University, 198504 St. Petersburg, Russia; vlar @astro.spbu.ru, HTH-home@yandex.ru.}
\altaffiltext{3}{Steward Observatory, University of Arizona, Tucson, AZ 85721-0065; psmith@as.arizona.edu.}
\altaffiltext{4}{Isaac Newton Institute of Chile, St. Petersburg Branch, St. Petersburg, Russia.}
\altaffiltext{5}{Multiple Mirror Telescope Observatory, University of Arizona, Tucson, AZ 85721-0065; gwilliams @mmto.org.}
\altaffiltext{6}{School of Physics and Astronomy, Cardiff University, Cardiff CF2 3YB, Wales, UK; walter.gear @astro.cf.ac.uk.}
\altaffiltext{7}{Lowell Observatory, Flagstaff, AZ 86001; buie@lowell.edu, taylor@lowell.edu, dunham@lowell.edu.}

\begin{abstract}
We explore the variability and cross-frequency correlation of the flux density and polarization of the blazar OJ287, using imaging at 43 GHz with the Very Long Baseline Array, as well as optical and near-infrared polarimetry.  The polarization and flux density in both the optical waveband and the 43 GHz compact core increased by a small amount in late 2005, and increased significantly along with the near-IR polarization and flux density over the course of 10 days in early 2006.  Furthermore, the values of the electric vector position angle (EVPA) at the three wavebands are similar.  At 43 GHz, the EVPA of the blazar core is perpendicular to the flow of the jet, while the EVPAs of emerging superluminal knots are aligned parallel to the jet axis.  The core polarization is that expected if shear aligns the magnetic field at the boundary between flows of disparate velocities within the jet.  Using variations in flux density, percentage polarization, and EVPA, we model the inner jet as a spine-sheath system.  The model jet contains a turbulent spine of half-width 1.2$\degr$ and maximum Lorentz factor of 16.5, a turbulent sheath with Lorentz factor of 5, and a boundary region of sheared field between the spine and sheath.  Transverse shocks propagating along the fast, turbulent spine can explain the superluminal knots.  The observed flux density and polarization variations are then compatible with changes in the direction of the inner jet caused by a temporary change in the position of the core if the spine contains wiggles owing to an instability.  In addition, we can explain a stable offset of optical and near-IR percentage polarization by a steepening of spectral index with frequency, as supported by the data.
\end{abstract}


\section{Introduction}
Blazars are powerful and variable sources of polarized radiation across a wide range of wavelengths.  However, establishing the location for this energetic and often dynamic emission within blazars is problematic.  In particular, efforts to determine the site of optical emission within the parsec-scale jet have been inhibited by the low spatial resolution available at this waveband.  Recently, new techniques have been developed to counter these difficulties.  Unresolved optical emission can be connected to a location in resolved Very Long Baseline Array (VLBA) images of the jet by comparing electric vector position angles (EVPA) and degrees of polarization, as well as variable total flux density across the two wavebands.  Studies utilizing this method have concluded that variable optical emission originates in the 43 GHz core \citep{lis00, gab06, dar07, jor07}.  Here, we extend this technique through time-variable correlations between optical and radio wavelengths in the blazar OJ287 (0851+202, $z=0.306$; \citeauthor{sit85} \citeyear{sit85}).  

OJ287 is a BL Lacertae object containing a one-sided jet with complex structure on parsec \citep{rob87} and kiloparsec \citep{per94} scales.  Previous studies have imaged the jet of OJ287 at milliarcsecond (mas) resolution at multiple radio wavelengths; the resultant maps contain a compact core with substantial linear polarization (P$\lesssim10\%$), and a tendency toward a $90^\circ$ difference between the EVPA of the core and the EVPA of the more extended jet \citep{rob87, gab96, ojh98}.  Historically, the inner jet has expelled components along westward paths, varying from position angle $-82\degr$ \citep{vic96} to $-122\degr$ \citep{tat04} with respect to the core.  \citet{tat04} model the change of inner jet position angle as cyclic, with period 11.6 years.  Such periodic variation, plus projection effects, agree with the tendency of OJ287 to establish moderately long-term trends, such as the observed position angle of $\sim-110\degr$ in the late 1980's and $\sim-90\degr$ in the mid 1990's \citep{kol92, gab96, tat99, jor01, jor05}.  In addition, \citet{jor05} noted the presence of several stationary knots in the jet of OJ287 with position angles ranging from $-90\degr$ to $-111\degr$, a possible lingering effect of the swinging inner jet. 

The inclusion of data at optical wavelengths has established a preliminary connection between the optical and radio EVPA in OJ287 \citep{rud78, gab94}.  Furthermore, \citet{kik88} and \citet{sil91} have measured rotations of EVPA at both wavebands, over scales varying from a month to a decade.  

We expand previous studies of OJ287 by combining resolved 43 GHz VLBA images of the parsec-scale jet with simultaneous polarimetric measurements at optical and near-IR frequencies over two intensive 10-day campaigns.  Our results show a clear connection between the polarized emission at optical and near-IR wavelengths and that associated with the 43 GHz parsec-scale core.  In addition, we find rapid, synchronous changes in radio, optical, and near-IR degree of polarization and total flux in early 2006, which we interpret as the consequence of a changing inner jet relative to the line of sight.

\section{Observations}
In our observational program, we utilized instruments at optical, near-infrared, and radio wavelengths to collect highly-sampled measurements of percentage polarization, EVPA, and flux density.  We obtained this information for 21 blazars during coordinated campaigns in 2005 October-November and 2006 March-April.  

To maximize the variability sampling, we observed at optical wavelengths using two facilities.  The majority of the data are from the 1.55 m Kuiper telescope at Steward Observatory, equipped with the SPOL spectropolarimeter \citep{sch92}.  We made ten spectropolarimetric observations of OJ287 between 2005 October 25 and November 3 and a further eight observations between 2006 March 27 and April 4.  For these 18 measurements, we utilized either a $3.0''$ or $4.1''$ slit width and a $10''$ spectral extraction aperture.  A diffraction grating with 600 lines per mm was used, producing a diffraction order of 400 to 800 nm with a dispersion scale of 0.4 nm pixel$^{-1}$.  We calculate the spectral resolution to be approximately 1.9 nm for a $3.0''$ slit width and 2.5 nm for a $4.1''$ slit width.  We binned the spectropolarimetry to derive polarization measurements within wavebands similar to those of the $B$ and $V$ Johnson filters and $R$ and $I$ Kron-Cousins filters.  This binning represents the continuum, as no line features were present in the spectra.  Photometric conditions existed on nine nights in 2005 and on two nights in 2006.  On these occasions, we measured the optical flux using either a $5.1\times 12''$, $7.6\times 12''$, or $7.6\times 14''$ extraction aperture during the 2005 campaign.  We employed a $7.6\times 12''$ extraction aperture in 2006.  We calibrated by comparison with field stars using magnitudes from \citet{smi85}.  A more detailed description of the procedural aspects of the spectropolarimetry can be found in \citet{smi03}.

We also performed optical observations at the Crimean Astrophysical Observatory on five nights between 2006 March 26 and 2006 April 5.  We took measurements on the AZT-8 70 cm telescope equipped with an ST7-based photometer-polarimeter, in which Savart plates are aligned at 45$\degr$ to one another.  The orientation of the plates thus produces a measurement of either the Stokes $Q$ or $U$ parameter.  Up to five images in each position of the plates allowed co-addition to improve the signal-to-noise ratio.  We obtained both polarimetric and photometric data in the $R$ waveband on each night using aperture photometry, and calibrated using standard field stars.  We also include in this study monitoring at this observatory outside the 2006 campaign period. OJ287 was not observed from the Crimean Astrophysical Observatory during the 2005 campaign.  All optical measurements are corrected for statistical bias \citep{war74}.

We observed in the near-infrared H-band (1.65 $\micron$) at Lowell Observatory's Perkins 1.8 m telescope.  Using the Mimir instrument \citep{cle07}, we obtained polarimetric and photometric measurements twice during the 2005 campaign and on 3 nights during 2006.  Polarimetry involved imaging OJ287 through a rotating half-wave plate and a wire-grid analyzer, with 32 images obtained over a full $360\degr$ in position angle of the wave plate.  Images were calibrated using a bad pixel mask, dark current and bias count subtraction, flat-field correction, a correction for the non-linear response of the detector, and EVPA calibration based on observation of polarimetric standards.  In addition, we performed differential photometry relative to nearby IR standard stars.

We observed OJ287 with the VLBA at 43 GHz on three dates spaced approximately equally across the 10 days of each campaign: 2005 October 24, October 28, and November 2, and 2006 March 27, March 31, and April 5.  Correlation of the data collected occurred at the Array Operations Center of the National Radio Astronomy Observatory (NRAO) in Socorro, New Mexico.  We calibrated the {\it uv} data using the Astronomical Image Processing System (AIPS) software from NRAO, and then constructed and analyzed the images with the DIFMAP software package \citep{she97}.  See \citet{jor05} for a complete discussion of our data reduction methods.  The EVPAs in the images were calibrated by summing the total and polarized emission across images of a number of objects, and comparing these integrated polarization characteristics with those supplied by NRAO from measurements made with the Very Large Array (VLA) at dates as close as possible to our epochs of observation.  The objects used, with the average separation in time in parentheses, were 0923+392 (4C +39.25) (6 d), BL Lac (3 d), and OJ287 (6 d) for the 2005 campaign, and 3C 273 (7 d), 3C 279 (7 d), OJ287 (7 d), and B2 1156+295 (4C +29.45) (7 d) during 2006.  The variance in the EVPA calibration, as determined using the integrated VLA data, was used to estimate the uncertainty in the calibration.  To check the validity of our EVPA calibration, we examine the change in instrumental D-terms between epochs, as well as the variance in EVPA of jet components with known values.  We conclude from this examination that the EVPA calibration is correct to within $\pm5\degr$ in 2005 and to within $\pm6\degr$ in 2006.  We apply a $30\degr$ compensation for Faraday rotation to the core region of OJ287, as derived by \citet{jor07}; unless otherwise noted, all EVPAs quoted have been corrected for this Faraday rotation.  To determine the motion of components within the jet, we use supplementary 43 GHz images of OJ287 observed with the VLBA at four epochs - 2006 May 16, June 17, August 10, and October 5.  These supplementary images are part of a monthly monitoring program of a sample of blazars, and have $uv$-coverage comparable to that obtained during the 2005 and 2006 campaigns.

\section{Analysis}
\label{Analysis}
\begin{figure}
\epsscale{0.72}
\plotone{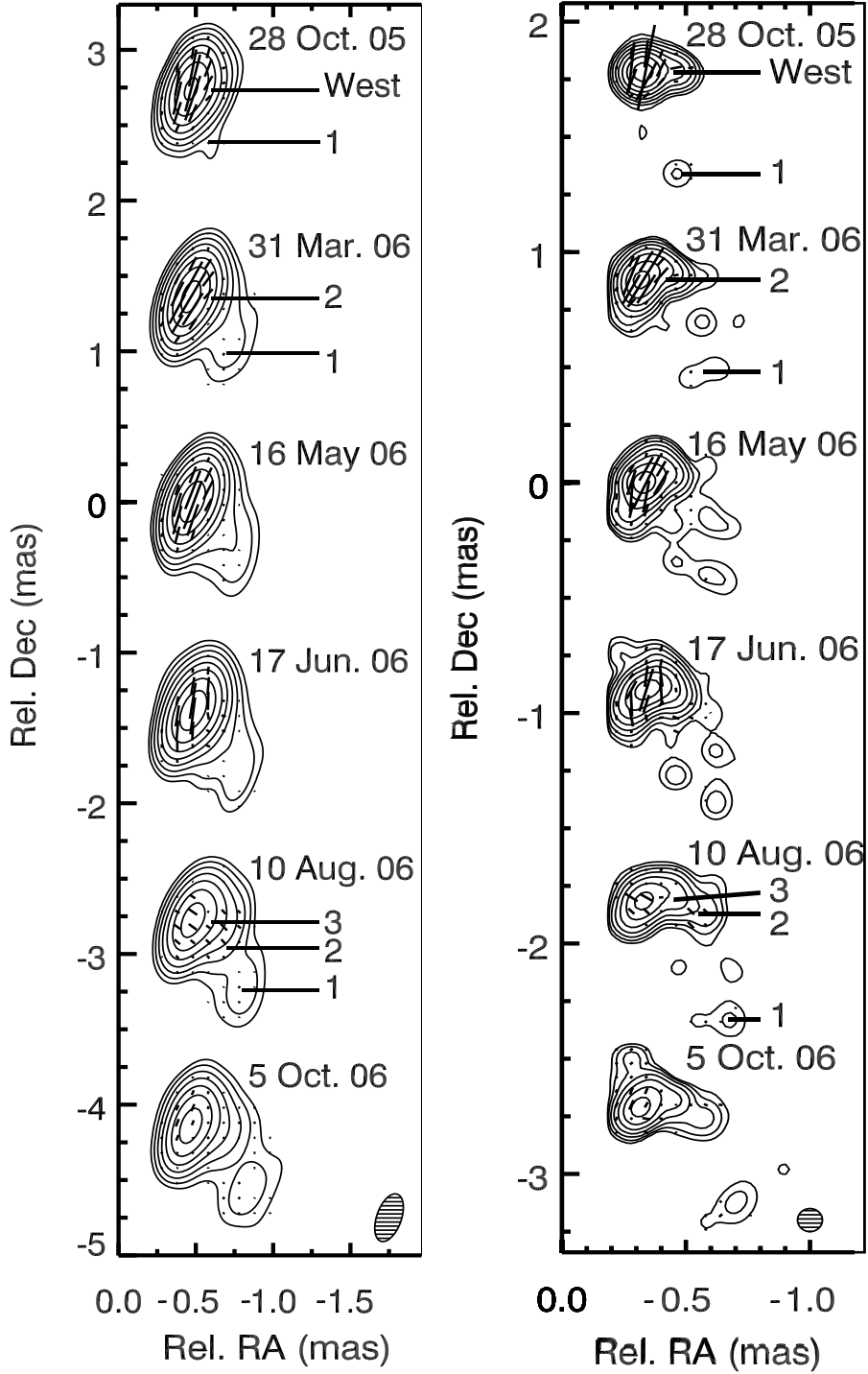}
\caption{43 GHz VLBA images of OJ287 at 6 epochs of observation (dates listed).  The bars represent polarized intensity, and the position angle of the bar corresponds to the EVPA.  Neither Faraday correction of the core EVPA nor statistical bias correction of the polarized intensity has been applied to this display of the images.  The contour levels are 0.5, 1, 2, 4, 8, 16, 32, and 64$\%$ of the peak total intensity. {\it Left:} The restoring beam (shown in the lower right corner) has dimensions of 0.33 mas $\times$ 0.16 mas at an angle of $-18.7\degr$. {\it Right:} The high-resolution restoring beam (shown in the lower right corner) has a circular diameter of 0.1 mas, and is representative of the resolution of the longest baseline. 
\label{figa}}
\end{figure}

The jet of OJ287 consists of both superluminally and subluminally moving components that extend from the core along a position angle (PA) that varies among the components.  We derive the total flux density, polarized flux density, position, and size of these components by fitting each $I$, $Q$, and $U$ map with circular Gaussian components using the task MODELFIT in Difmap.  Components are fit in order of peak flux density, from largest to smallest.  A detailed description of the fitting technique can be found in \citet{jor05}.  Table 1 lists the fitted parameters for each component - position relative to the core, PA, total flux density, degree of polarization, and EVPA - as well as the reduced chi-square statistic for the cumulative model after adding each component, starting with the core.  In Figure \ref{figa}, we show a VLBA image for each of the two campaigns, as well as a VLBA image from each of the four supplementary epochs,  with positions of three components marked.  In the right panel, we utilize a small circular Gaussian restoring beam of 0.1 mas FWHM diameter, corresponding to the resolution of the longest baselines, to demonstrate the presence of components close to the core. 

\begin{figure}
\epsscale{1.00}
\plotone{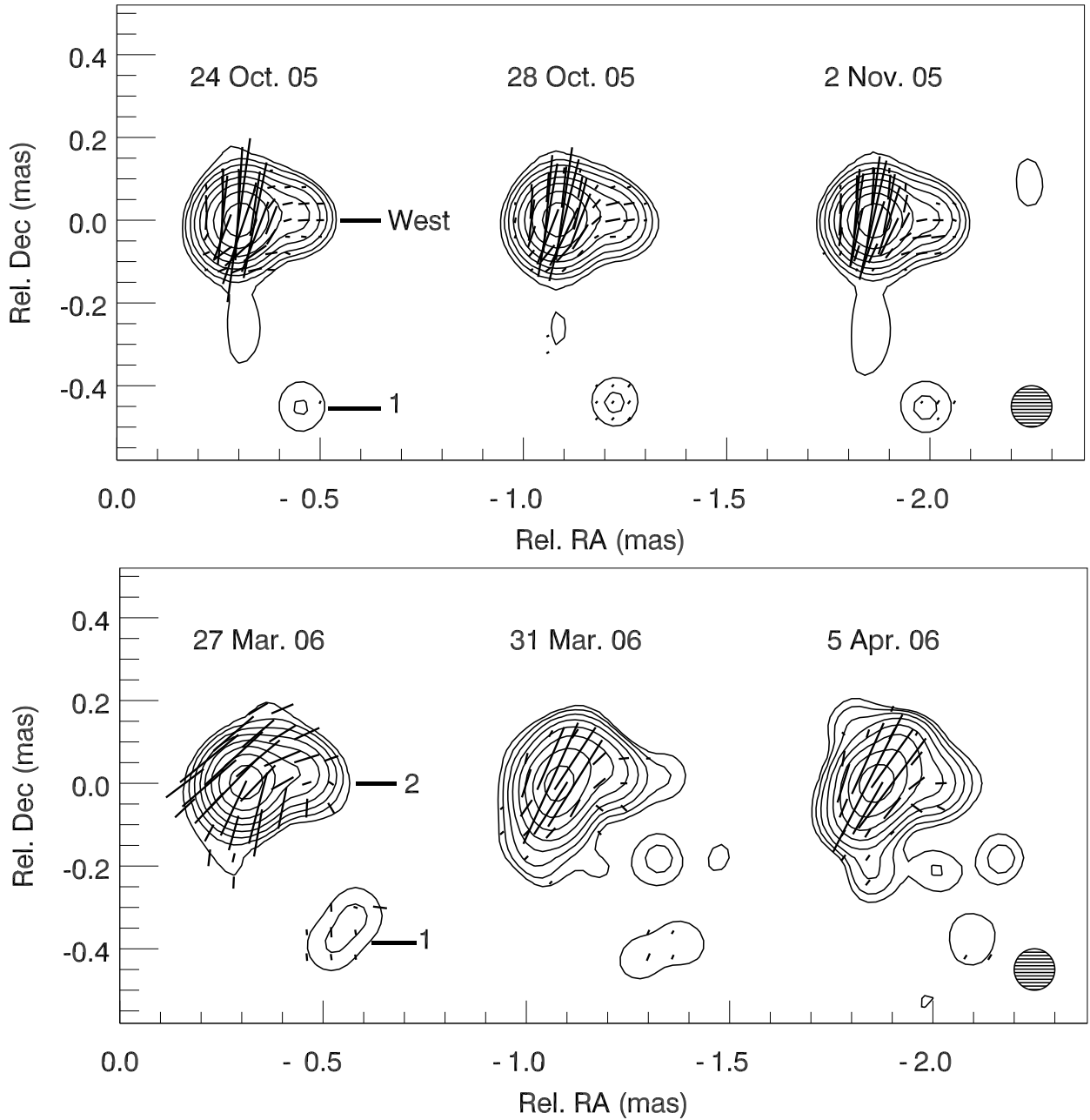}
\caption{43 GHz VLBA images of OJ287 at 3 epochs of observation for ({\it Top}) the 2005 campaign and ({\it Bottom}) the 2006 campaign (dates listed).  The contour levels are 0.5, 1, 2, 4, 8, 16, 32, and 64$\%$ of the peak total intensity.  The restoring beam (shown in the lower right corner) has a circular diameter of 0.1 mas, and is representative of the resolution of the longest baseline.
\label{figzoom}}
\end{figure}

In October 2005, the submilliarcsecond structure of OJ287 included the emergence of a new feature.  This component is labeled in Table 1 and all figures as the ``West Component''.  In Figure \ref{figzoom} ({\it top}), we display the three high-resolution images of this component from the 2005 campaign.  These images reveal the presence of a strong core with flux density $2.2\pm0.2$ Jy, EVPA $\chi=19\degr\pm4\degr$ (after correction for Faraday rotation), and degree of polarization $P=12.1\pm0.5\%$, and a component with flux density $0.8\pm0.2$ Jy to the west (PA $-74\pm14\degr$) with $\chi=-79\pm10\degr$ and $P=6.9\pm0.8\%$.  We do not detect this component during the next campaign (2006 March).

\begin{figure}
\epsscale{1.00}
\plotone{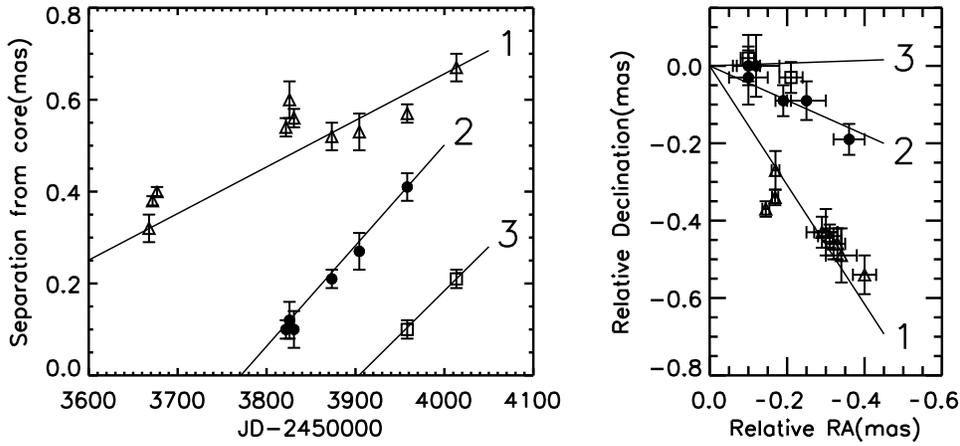}
\caption{{\it Left:} Separation in milliarcseconds relative to the core of three components, with one point plotted for each epoch at which the component appears on the respective VLBA image.  The triangles (1) correspond to the component with birth date $\sim$2004 March, the filled circles (2) to the component with birth date $\sim$2006 February, and the open squares (3) to the component with birth date $\sim$2006 May.  Robust least-squares fitting to the separation of each component is shown with solid lines.  Julian date 2453700.5 corresponds to 2005 November 26.  {\it Right:} Map of the locations of the three components.  Robust least-squares fitting to the position angle of each component is shown with solid lines.  The motion is outward from the core, located at position (0,0).  
\label{fig2}}
\end{figure}

During the 2006 campaign, a component (labeled ``2" in Figures \ref{figa}, \ref{figzoom}, and \ref{fig2}) separated from the core to the west at PA = $-97\pm11\degr$, with long-term motion along PA = $-114\pm4\degr$.  We first imaged this knot during the 2006 campaign, and subsequently in 2006 May, June, and August.  The emergence of the component in the three epochs of observation of the 2006 campaign is shown in high resolution in Figure \ref{figzoom} ({\it bottom}).  During the campaign, the new knot was positioned ~0.1 mas to the west of the core, with a flux density of $1.8\pm0.5$ Jy and $\chi=77\pm4\degr$, while the core had $\chi=-1\pm4\degr$, after correction for Faraday rotation. The knot retained a steady value of $\chi=66\pm11\degr$ between March 31 and August 10.  Its flux density dropped steeply and steadily, as found for three previous knots in OJ287 by \citet{jor05}, to a value of 0.14 Jy on 2006 August 10.  

A plot of separation from the core versus time (Figure \ref{fig2}, {\it left}) reveals that the proper motion of component 2 is approximately constant throughout its lifetime at 0.8 $\pm$ 0.1 mas yr$^{-1}$.  This corresponds to an apparent speed of 12$\pm1$c, and a ``birth'' (coincidence with the core) on 2006 February 4 $\pm$ 27 days.\footnote[1]{We use the cosmological parameters H$_{0}$ = 72 km/s/Mpc, $\Omega_{m} = 0.27$, and $\Omega_{\Lambda} = 0.73$ to calculate all speeds and length scales.}

In late 2006, we observed a component moving along a more westward long-term path, $-88\pm14\degr$ (labeled ``3" in Figures \ref{figa}, \ref{figzoom}, and \ref{fig2}).  From the 43 GHz images of 2006 August 10 and October 5, we derive an angular velocity of 0.7$\pm$0.2 mas yr$^{-1}$, or an apparent speed of 10$\pm3$c.  We then calculate a birth date for this feature of 2006 June 17 $\pm$ 33 days.  Within the two dates of observation, the flux density of this component decreases by a substantial amount, as can be seen in Table 1.  

In addition, we map the path of an older, slower component along a position angle of $-147\pm5\degr$ (labeled ``1" in Figures \ref{figa}, \ref{figzoom}, and \ref{fig2}).  Component 1 appeared in the images of both the 2005 and 2006 campaigns, as well as in 2006 May, June, August, and October.  We observe it to have $\chi=-18\pm15\degr$ throughout its lifetime.  Intermittent excursions in $\chi$ , as well as substantial and significant excursions in $P$ and flux density for this component suggest interaction with the surrounding medium.  We calculate an angular velocity of 0.37$\pm$0.04 mas yr$^{-1}$, or an apparent speed of 5.3$\pm0.6$c, which leads to a birth date of 2004 December 13 $\pm$ 77 days.  Figure \ref{fig2} ({\it right}) shows the paths of the components born in 2004 December, 2006 February, and 2006 June.

\begin{figure}
\epsscale{1.11}
\plottwo{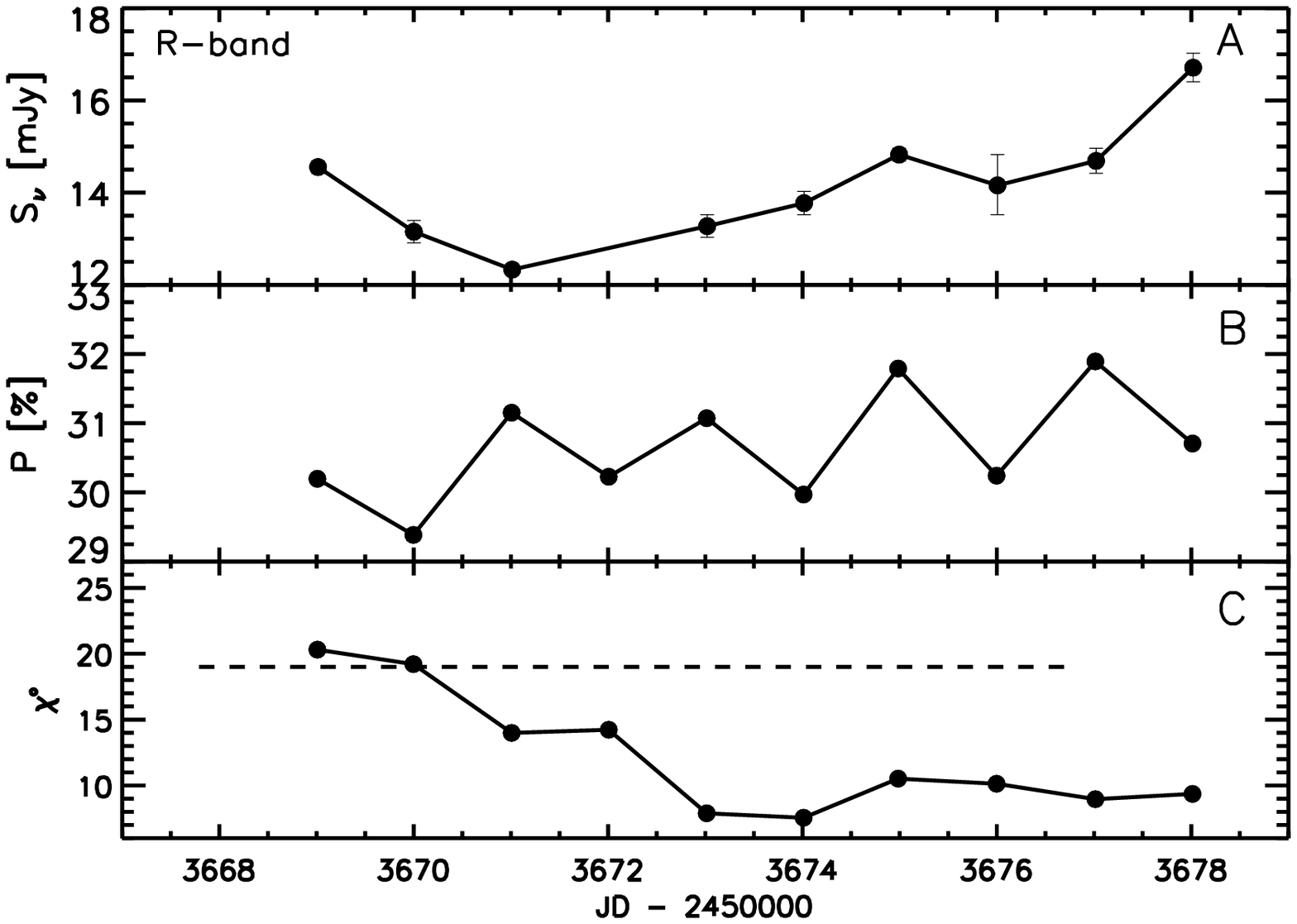}{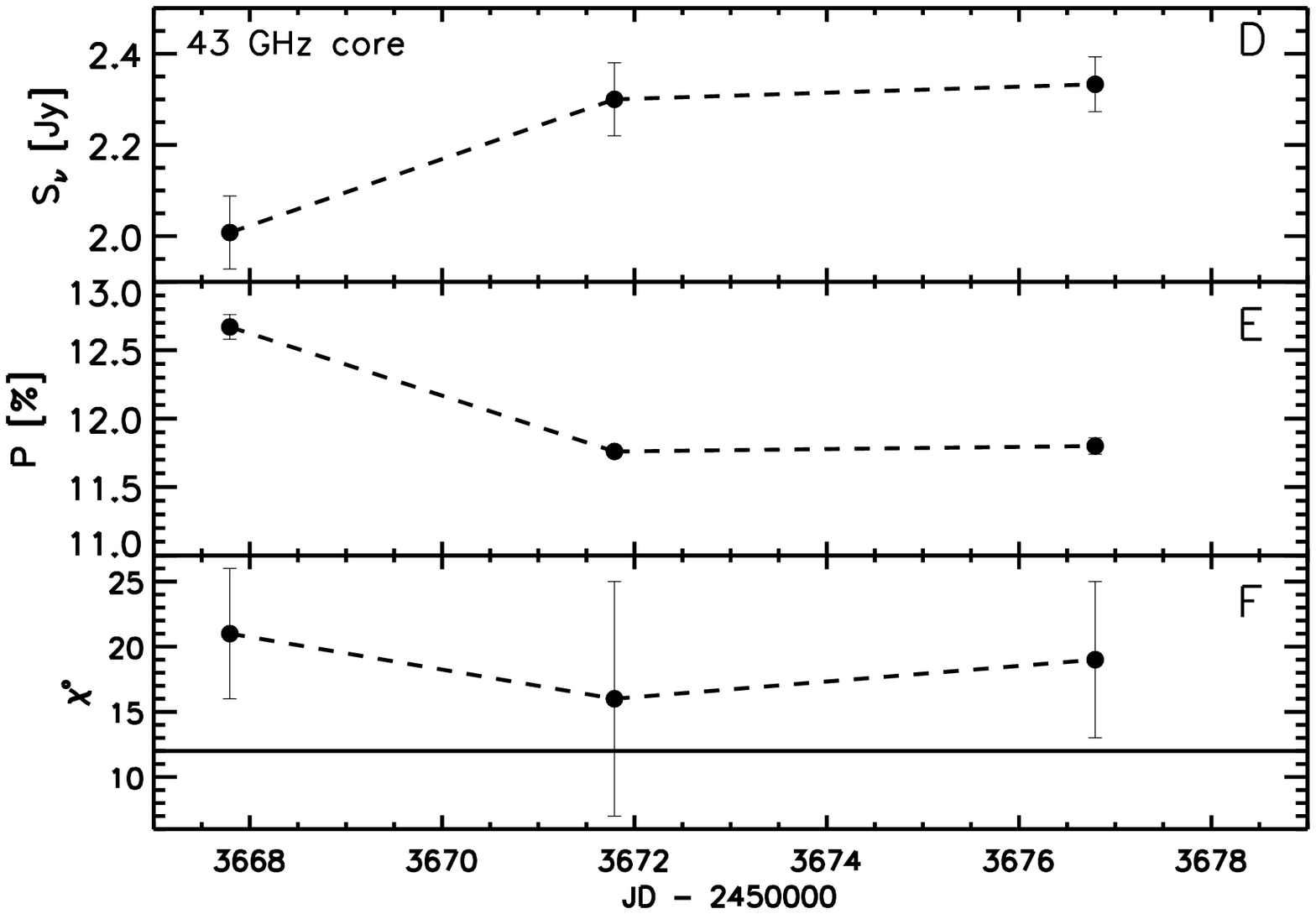}
\caption{Data from the 2005 campaign for OJ287.  {\it Left:} Data in $R$-band as a function of time showing flux density {\it (A)} corrected for reddening \citep{sch98}, percent polarization {\it (B)}, and $\chi$ {\it (C)}.  The dashed line represents the average $\chi$(43 GHz core).  {\it Right:} Data  for the 43 GHz core displayed in a similar manner as the optical data.  In all cases, $\chi$(core) has been corrected by 30$\degr$ to account for Faraday rotation.  The solid line represents the average $\chi$($R$).  Julian date 2453668.5 corresponds to 2005 October 25.
\label{fig3}}
\end{figure}

\begin{figure}
\epsscale{1.11}
\plottwo{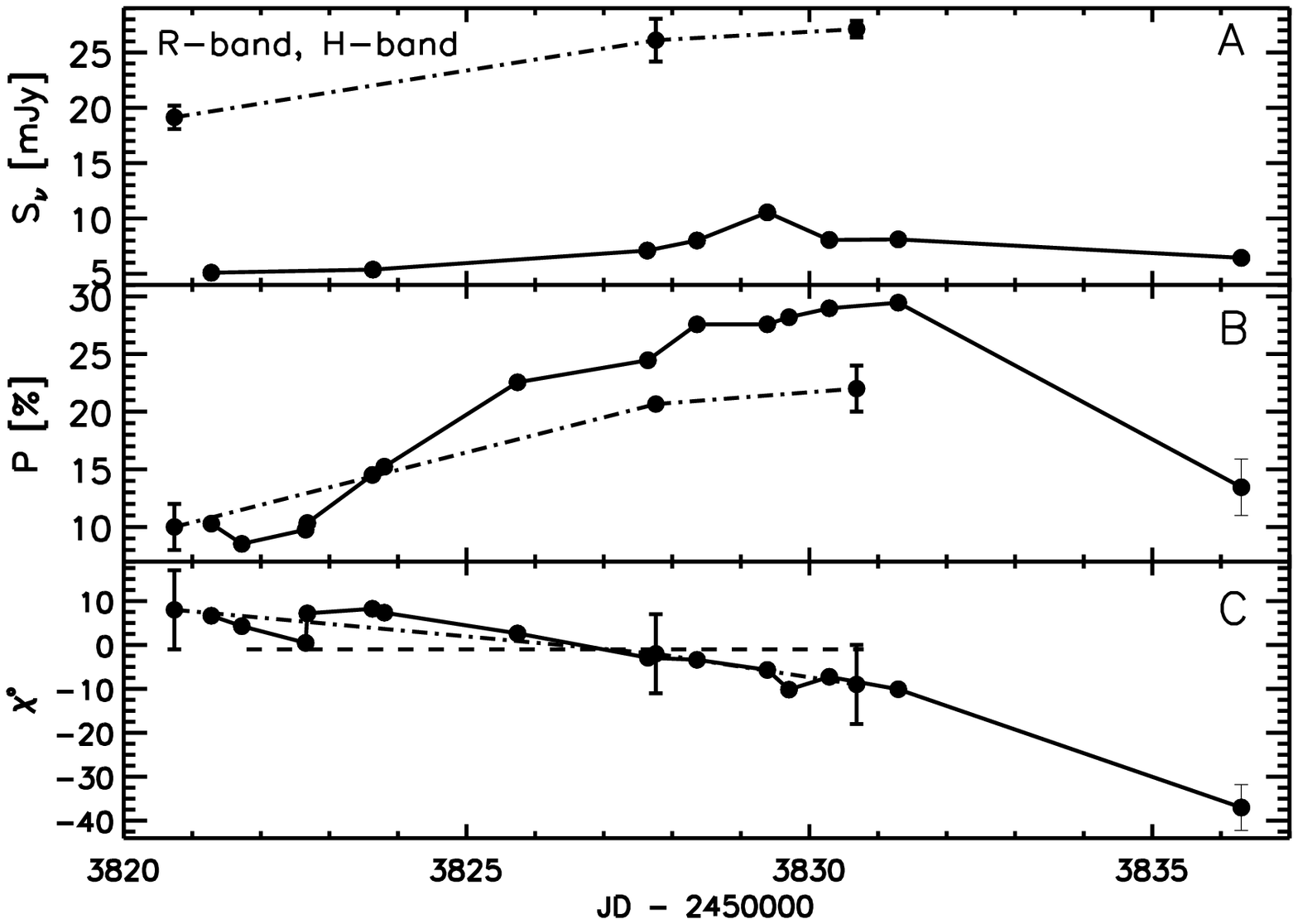}{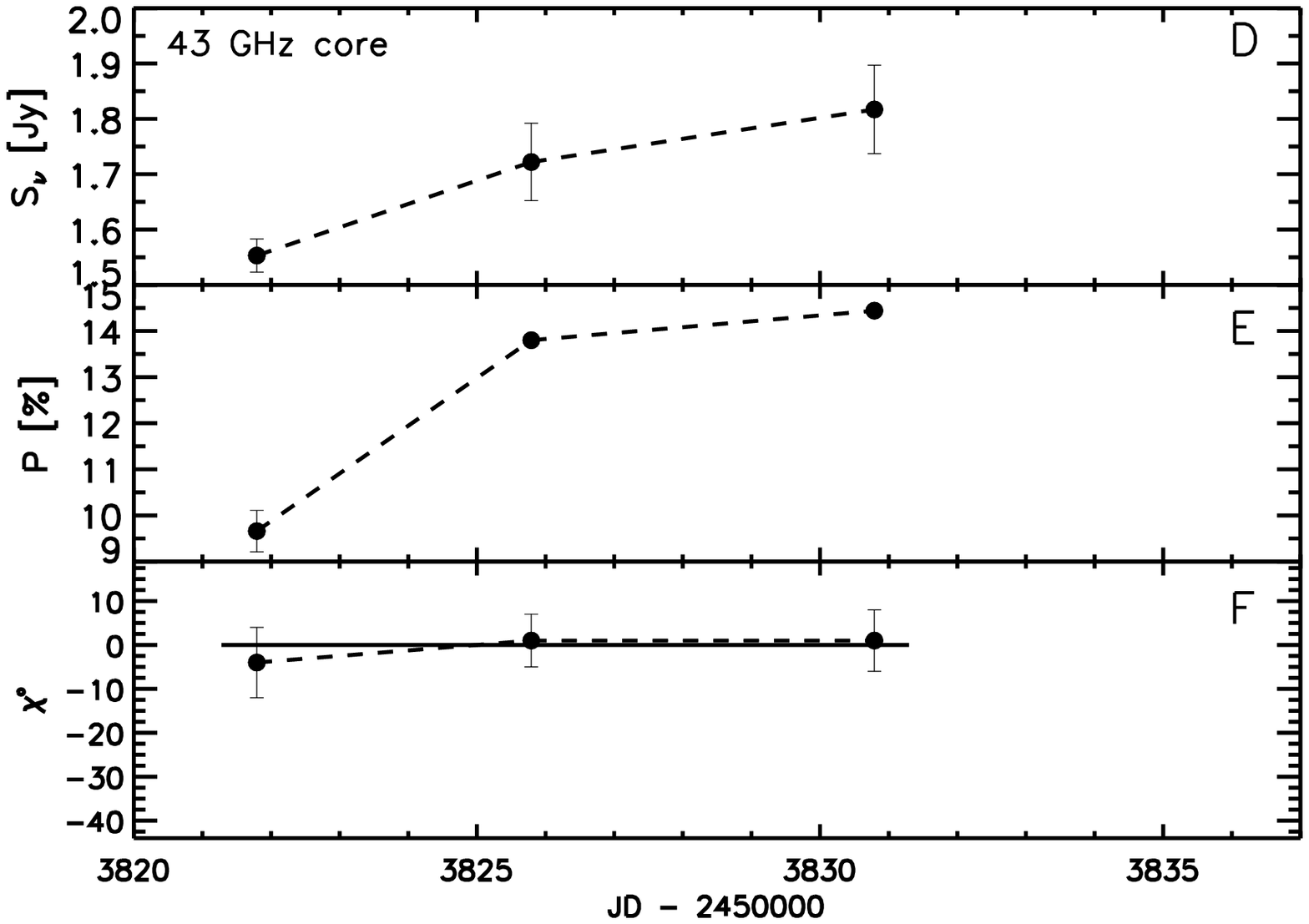}
\caption{Measurements from the 2006 campaign for OJ287 shown in the same manner as in Figure 4, except for the addition of $H$-band measurements (in {\it (A)}, {\it (B)}, and {\it (C)}) represented by a dash-dotted line.  Julian date 2453820.5 corresponds to 2006 March 26.
\label{fig4}}
\end{figure}

Figure \ref{fig3} shows both $\chi$(optical) and $\chi$(43 GHz core) during the 2005 campaign.  We find a short rotation (from $20\degr$ to $8\degr$), then plateau, of $\chi$(optical).  The average $\chi$($R$-band) is $12\pm5\degr$, the average $\chi$($H$-band) is $8\pm6\degr$, and the average $\chi$(43 GHz core) is $19\pm4\degr$, comparable values.  Within the 2006 campaign (JD 2453820 - JD 2453832), shown in Figure \ref{fig4}, we observe a steady rotation of $-2\degr$ day$^{-1}$ in both $\chi$(optical) and $\chi$(H-band), with average values of $\chi$($R$-band)$=0\pm7\degr$ and $\chi$($H$-band)$=-1\pm9\degr$, while the 43 GHz core value is constant at $\chi$(43 GHz core)$=-1\pm4\degr$.  Rotation in the latter is not apparent, perhaps due to the rather high uncertainties, but $\chi$(43 GHz core) agrees with the average $\chi$(optical) and $\chi$(H-band) within the errors throughout the period of observation. 

\begin{figure}
\epsscale{1.00}
\plotone{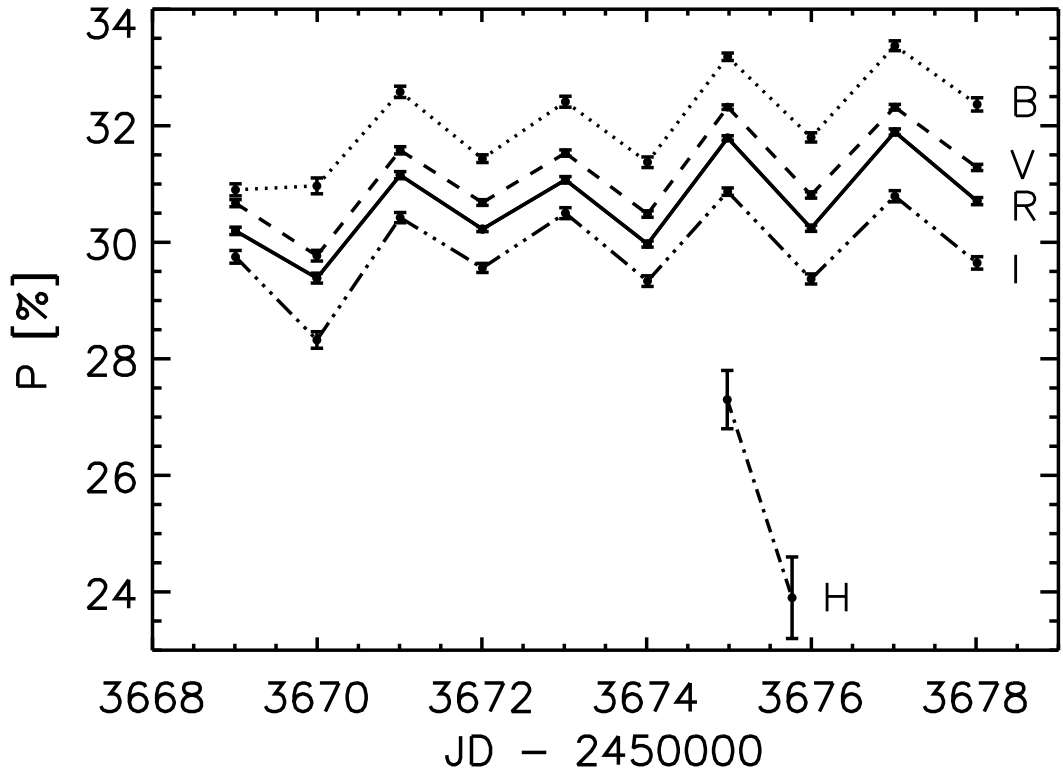}
\caption{Multiwavelength plot of polarization variations with time during the 2005 campaign.  $B$ waveband represented by dotted line, $V$ by dashed line, $R$ by solid line, $I$ by dash-triple dotted line, and $H$ by dash-dotted line.  Julian date 2453668.5 corresponds to 2005 October 25.
\label{fig5}}
\end{figure}

The degrees of polarization in the optical and radio core are nearly steady throughout the 2005 campaign, as shown in Figure \ref{fig3}; with $P$($R$) increasing by only $\sim 0.1\%$ day$^{-1}$.  We measure the mean radio core polarization to be $12.1\pm0.5\%$, the mean $H$-band value to be $26\pm2\%$, and the mean $R$-band value to be $30.7\pm0.8\%$, with a consistent ratio $P$($R$)/$P$(43 GHz core) = 2.5.  We also note that the optical polarization appears to oscillate with a period of roughly 2 days that is observed at all optical wavelengths (see Figure \ref{fig5}).  We observed stability in 5 polarized standard stars during these observations, as well as polarization $<0.1\%$ for 2 unpolarized standards, leading us to conclude that this oscillation is not a systematic, instrumental effect.  In addition, we note that this behavior is not observed in any other blazar during the campaign.  Furthermore, \citet{smi87} find fluctuations in the degree of polarization of OJ287 of a similar period.  We conclude, due to the repeating nature of this behavior, that the effect is intrinsic to the blazar, and not an artifact of propagation through intervening media.  However, we do not have 43 GHz sampling on a similar timescale, so that the data are insufficient to draw any conclusions as to the structural basis for this variation.   

The flux density underwent a modest change during the 2005 campaign.  There is a $36\%$ increase in $R$-band flux density from 12.3 to 16.7 mJy and a less dramatic increase in 43 GHz core flux density from 2.0 to 2.3 Jy.  We fit the optical spectra resulting from the spectropolarimetry, corrected for Galactic extinction using \citet{sch98}, with a power law and find an optical spectral index, steady over the campaign, of $\alpha_{opt} = -1.28\pm0.05$, where the total flux density $F_\nu \propto \nu^\alpha$.    

\begin{figure}
\epsscale{1.00}
\plotone{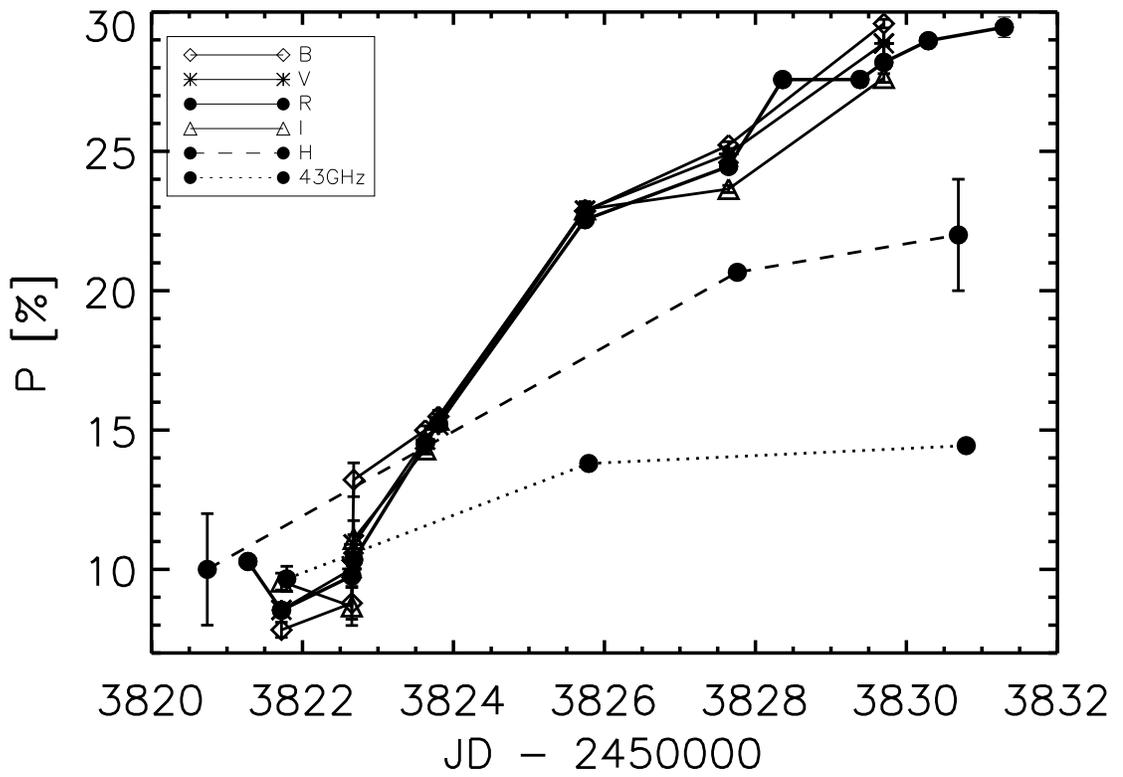}
\caption{Multiwavelength plot of polarization variations during the 2006 campaign.  Julian date 2453820.5 corresponds to 2006 March 26.
\label{fig6}}
\end{figure}

During the 2006 campaign, the optical, infrared, and radio core polarizations increased continuously (see Figure \ref{fig4}), with no significant time delay between frequencies.  The optical $R$-band polarization began at $8.5\%$ on 2006 March 27 and grew to $29.5\%$ by 2006 April 5.  A least-squares fit to $P$ versus time gives a growth rate of $2.30\pm0.01\%$ day$^{-1}$.  The infrared polarization increased from $10\%$ to $22\%$, and $P$(43 GHz core) from $9.7\%$ to $14.4\%$.  We note that polarization at all wavelengths is approximately the same near 2006 March 27, $\sim10\%$.  After that date, the polarization increases at all wavebands, but the increase is greater at higher frequencies (see Figure \ref{fig6}).  The dominance of wavelength-dependent polarization effects grows with the percentage polarization.

In addition, the optical, near-IR, and 43 GHz flux densities increased during the 2006 campaign, as shown in Figure \ref{fig4}.  In $R$-band, we observed a steady doubling in flux density from 5.2 mJy on March 26 to 10.6 mJy on April 3 before it declined on April 4 and 5.  The infrared $H$-band emission increased from 19.1 mJy on March 26 to 27.1 mJy on April 5.  In the 43 GHz core, the flux density rose in a similar fashion, from 1.55 Jy on March 27 to 1.82 Jy on April 5.  From the optical measurements, which had a higher frequency of sampling than either those at 43 GHz or $H$-band, we conclude that the local maximum occurred on April 3 (JD 2453829).  We also observed the optical spectral index to have a stable value of $\alpha_{opt} = -1.3\pm0.1$, which agrees within the measurement uncertainty with the index derived from the 2005 data.  Combining the 5 $H$-band flux measurements over both campaigns, we calculate an average spectral index between $H$-band and $I$-band of $\alpha_{H-I} = -0.9\pm0.1$.

\section{Discussion}
\subsection{Multiwaveband Polarization Characteristics}
Since the dramatic increase in the degree of polarization of OJ287 during the 2006 campaign occurs both at optical/near-infrared wavelengths and in the emission from the 43 GHz core, it indicates that the regions of emission are at least partially cospatial.  Further evidence for this is provided by the close correspondence of $\chi$(optical/IR) and $\chi$(radio core) during both the 2005 and 2006 campaigns.  {\it We conclude that the variable optical/near-infrared emission originates from the core region of the parsec-scale jet.}  

The polarization vector in the core of OJ287 is oriented roughly perpendicular to the flow of the jet.  Under the standard relativistic synchrotron model for production of optically thin polarized emission - wherein $\chi$ is transverse to the magnetic field lines - the orientation of $\chi$ in OJ287 reveals that the mean magnetic field in the core is aligned parallel to the flow of the jet.  To more precisely map the orientation of the magnetic field in the jet, we determine the closest observed epoch to the ejection of each component, and note $\chi$(core) at each of these epochs.  We find a consistent difference of $99\pm13\degr$ between $\chi$(core) at the time of ejection and the subsequent path of the emerging components.  Therefore, the observed magnetic field in the core is approximately aligned with the axis of the jet at any given epoch, parallel to the direction of flow.  A region of velocity shear in the core is the simplest explanation for this observed field geometry.  The shear could be produced when fast plasma near the jet axis flows past slower material closer to the boundary, resulting in a gradient of velocities that stretches and aligns the magnetic field, which may have originally been randomly oriented in turbulent cells, along the direction of flow.

We also investigate the possibility that multiple radio components contribute to the change in polarization, and conclude that this is not a feasible solution.  We observe a distinct stability of the orthogonality between $\chi$(core) and the jet axis, even for significant changes in the projected PA of the jet.  Combined with the constant agreement between $\chi$(core) and $\chi$(R), we conclude that the changes in $\chi$ and $P$ are consistent with intrinsic changes in the viewing angle of the jet, which causes changes in the observed angle of aligned magnetic field.  In addition, throughout the period of increasing polarization, we observe no variation in optical spectral index with time or frequency, nor do we observe any evidence for wavelength-dependent $\chi$.  These are two tendencies that are commonly observed in the case of multiple-component flares in OJ287 \citep{hol84} and other blazars \citep{bri86}. 

\subsection{Model for Polarization Behavior}
During the 2006 observations of OJ287, we see the optical flux density increase, then decrease, while the percentage polarization continues to grow (Figure \ref{fig4}).  To explain this behavior, we create a structural model for the inner jet as a combination of three elements.  The first element is a large, relatively slow turbulent sheath, which surrounds a thinner, faster, turbulent spine along the axis of the inner jet.  The transition region between the two zones produces the final element, with the magnetic field there aligned longitudinally by shear.  We use $\Gamma=16.5$ as the average Lorentz factor of the spine, derived from observations of the fast jet components of OJ287 in the detailed VLBA study of \citet{jor05}.  The Lorentz factor of the bulk jet flow is calculated from the observed pattern speeds of components - which cannot exceed $\Gamma$ - and the concurrent jet viewing angle, and is therefore greater than or equal to the fastest observed component speed.  In addition, we set the half-opening angle of the spine to be within the range $0.8\pm0.4\degr$, also derived for OJ287 in \citet{jor05}.  The average Lorentz factor of the sheath is stepped through a series of values to find the best fit, and the boundary region is assigned a Lorentz factor that is the arithmetic mean between $\Gamma_{\rm spine}$ and $\Gamma_{\rm sheath}$.  As the sheath is a large structure, we hold its viewing angle steady, using the value of $3.2\degr$ derived in \citet{jor05}.  We then allow the viewing angle of the small spine and boundary region to vary with time.  As the viewing angle of a spine-boundary structure changes when the jet direction swings slightly, the effects on the Doppler beaming of the radiation from this region are dramatic, while balanced by the radiation of the larger, slower sheath.  This yields a combination of a strongly varying component superposed on one having a relatively steady flux density.  

We initially reproduce the $R$-band flux density and polarization variations with this model, as it is the waveband with the highest sampling during the campaign.  We assign to all emission elements the average observed optical spectral index of $-1.29$.  To model the degree of polarization and flux density of the optical emission, we utilize the equations for total flux density from optically thin sources derived by \citet{caw06}, based on previous work by \citet{hug85}. The models of \citet{caw06}, created to describe synchrotron emission from conically shocked plasma, provide equations for the total flux density from regions of both aligned and turbulent magnetic field.  The proportionalities, adapted to a variable $\alpha$, are as follows:

\begin{equation}
F_{||} \varpropto (B_{||} \sin \theta_{||}^{*})^{1-\alpha} \nu^{\alpha} \delta_{||}^{2-\alpha}
\end{equation}
and
\begin{equation}
F_{\rm turbulent} \varpropto B_{\rm turb}^{1-\alpha} \frac{1}{\kappa^{2}} (2-(\sin \theta_{\rm turb}^{*})^{1-\alpha}(1-\kappa^{2})) \nu^{\alpha} \delta_{\rm turb}^{2-\alpha} \ .
\end{equation}

In the above equations, $\delta$ is the Doppler factor, which is a function of viewing angle $\theta$, jet speed $\beta$, and Lorentz factor $\Gamma$: $\delta = [\Gamma(1-\beta \sin \theta)]^{-1}$.  The compression factor due to shocking in the region is described by factor $\kappa$ and $B_{||}$ and $B_{\rm turb}$ are, respectively, the parallel and turbulent components of the magnetic field.  The influence of the magnetic field on flux density is described with the angle $\theta^{*}$, the angle between the aligned magnetic field and the line of sight, corrected for relativistic aberration.  We adapt these formulae for our usage by  instituting a lack of compression (i.e., no shock, $\kappa = 1$) and by removing the $\nu$ factor, as we are modeling a single waveband.  Applying our observed optical spectral index to these formulae, we find the following proportionalities for flux density in the $R$ waveband:

\begin{equation}
F_{\rm R, spine} \varpropto \delta_{\rm spine}^{3.29} B_{\rm turb, spine}^{2.29} \ ,
\end{equation}
\begin{equation}
F_{\rm R, transition} \varpropto \delta_{\rm transition}^{3.29} B_{||}^{2.29} (\sin \theta^{*})^{2.29} \ ,
\end{equation}
and
\begin{equation}
F_{\rm R, sheath} \varpropto \delta_{\rm sheath}^{3.29} B_{\rm turb, sheath}^{2.29} \ ,
\end{equation}
in which we approximate the parallel and turbulent components of the magnetic field to be roughly equal for the purpose of simplicity in the model.  This approximation is valid under the assumption that the magnetic field in the turbulent cells is aligned by shear, but not enhanced.  The equations for total flux density from the turbulent sheath and spine regions contain factors to account for the strength of the magnetic field and for the relativistic beaming effect.  The equation for total flux density from the transition region of aligned magnetic field contains an additional factor $(\sin \theta^{*})^{1-\alpha}$, reflecting the directionality of synchrotron emission, which peaks at a rest frame viewing angle perpendicular to the magnetic field.  

\begin{figure}
\epsscale{1.00}
\plotone{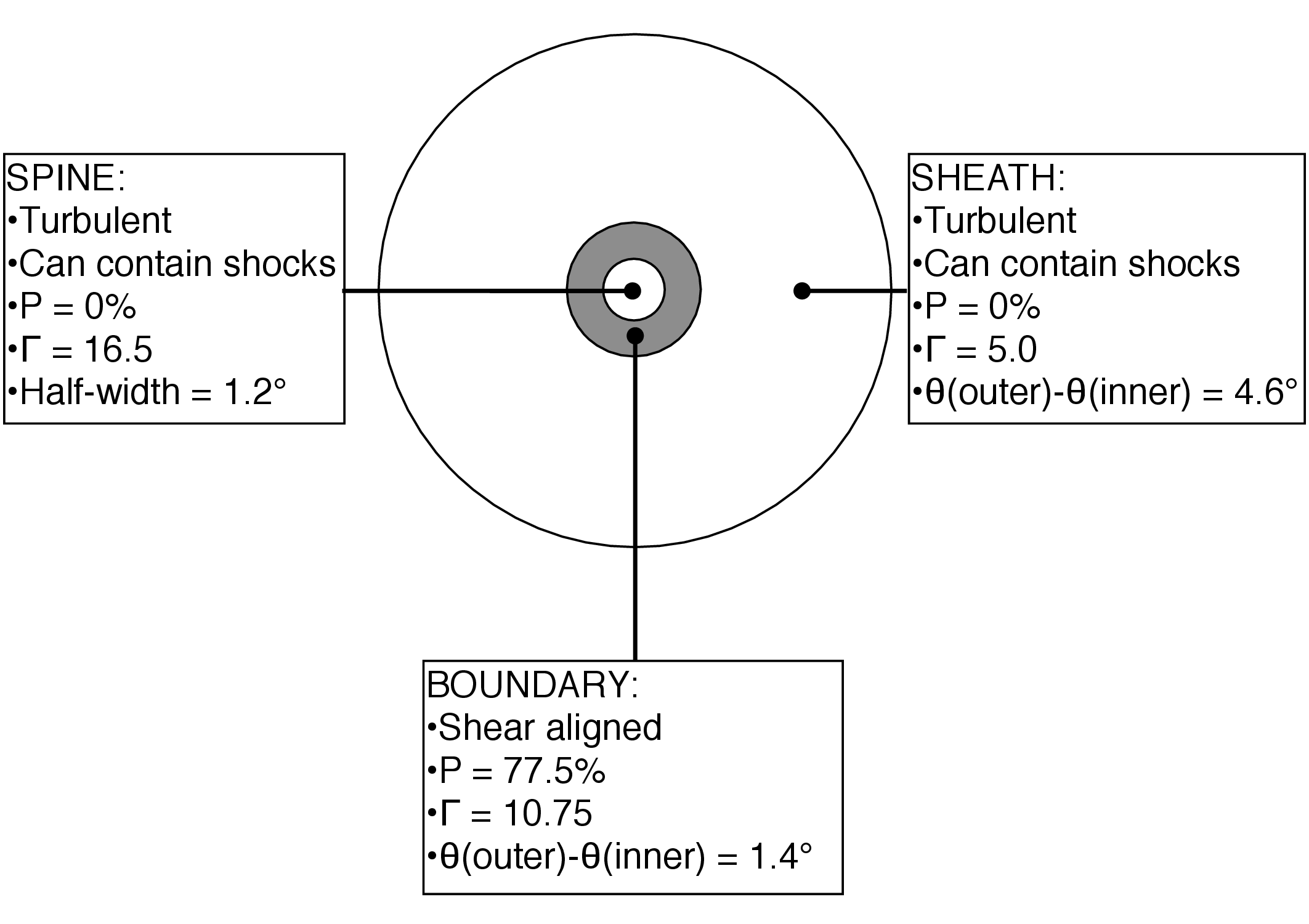}
\caption{Cross-sectional diagram of modeled jet in the vicinity of the core, with nominal quantities for the spine, sheath, and boundary regions listed.
\label{fig7}}
\end{figure}

\begin{figure}
\epsscale{1.00}
\plotone{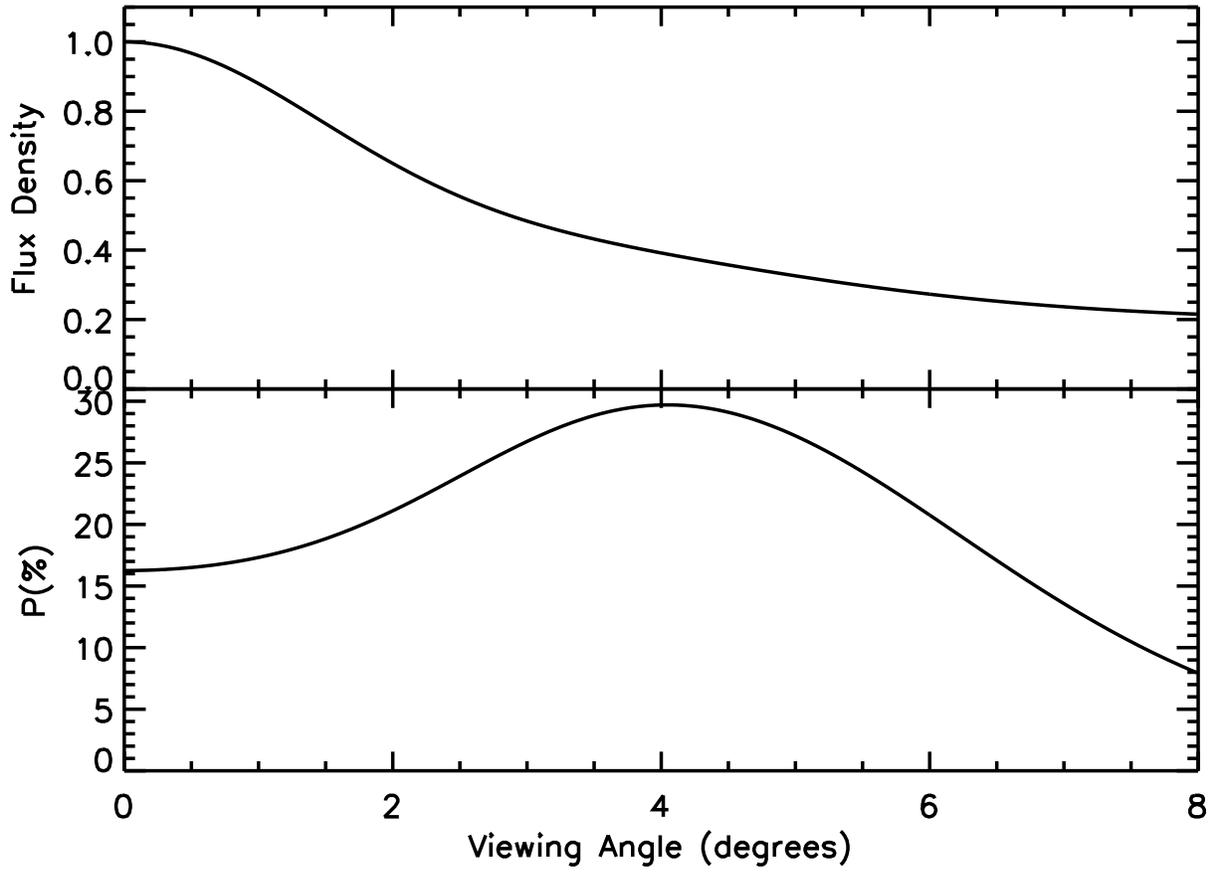}
\caption{Modeled flux density ({\it top}) and degree of polarization ({\it bottom}) as a function of viewing angle.  Flux density has been normalized to the peak.  Polarization peaks at a viewing angle of $4.05\degr$, and flux density at $0.0\degr$.
\label{fig8}}
\end{figure}

Utilizing these equations, we consider a circular cross-section of the jet.  See Figure \ref{fig7} for a cartoon of the model structure.  For each potential value of the viewing angle (calculated for 1000 steps between $0\degr$ and $8\degr$), we integrate the flux density across the spine and sheath, where the jet cross-section is separated into concentric rings with width $0.1\degr$.  We consider a range of values of the width of the sheath, the width of the spine \citep[within the errors stated above from][]{jor05}, and the average Lorentz factor of the sheath, to find the best model fit.  We calculate the degree of polarization by approximating that the turbulent sheath and chaotic magnetic field component in the spine have negligible polarization, and that the boundary region has the maximum polarization possible for synchrotron radiation, $77.5\%$ for a spectral index $\alpha=-1.29$.  We use this simplifying assumption to minimize the number of parameters, given our limited sample size.  Further monitoring of this blazar would allow the introduction of additional modeling factors, such as the level of magnetic field alignment.  We then manipulate the viewing angle of the spine and boundary region to simulate variations in flux density and polarization.  Figure \ref{fig8} displays the modeled quantities, with values discussed below.  

Two factors contribute to the changes in observed flux density with viewing angle.  The first is the Doppler factor, the decline of which decreases the flux density with increasing viewing angle.  In addition, the effect on the emission of the ordered component of the magnetic field, parallel to the jet axis, varies with the sine factor of equation (4).  As the de-aberrated viewing angle increases up to $90\degr$ - up to an observer's viewing angle in radians of $\sim1/\Gamma$ - the observed flux density from the ordered component increases.  These two multiplicative factors affect the observed emission of the boundary layer such that the flux density first grows, then declines, as the viewing angle increases while the spine changes direction.  The flux density of the turbulent spine is, however, only affected by the Doppler factor. The resultant combined flux density for all components peaks at a lower viewing angle than does the peak of total polarization.  This leaves an observable signature, in which the flux density of the combined system can grow to a maximum while the degree of polarization rises less steeply and peaks later in time (see Figure \ref{fig4}).

To model the observed flux density, we find the viewing angle required for an observed value of $P$($R$-band), then determine the modeled flux density corresponding to that viewing angle.  We compare the modeled flux density to the observed quantity to find the closest fit using our three variables.  The best fit is found for a spine of half-width $1.2\degr$, a sheath of width $4.6\degr$ (between the outer and inner boundaries), and a Lorentz factor in the sheath of 5.0.  Figure \ref{fig8} shows the modeled flux density and polarization graphs for this simulation.  We find that a decrease in viewing angle from $7.5\degr$ to $3.2\degr$ can create a simultaneous increase in flux and polarization, as observed during the 2006 campaign.  

\begin{figure}
\epsscale{1.00}
\plotone{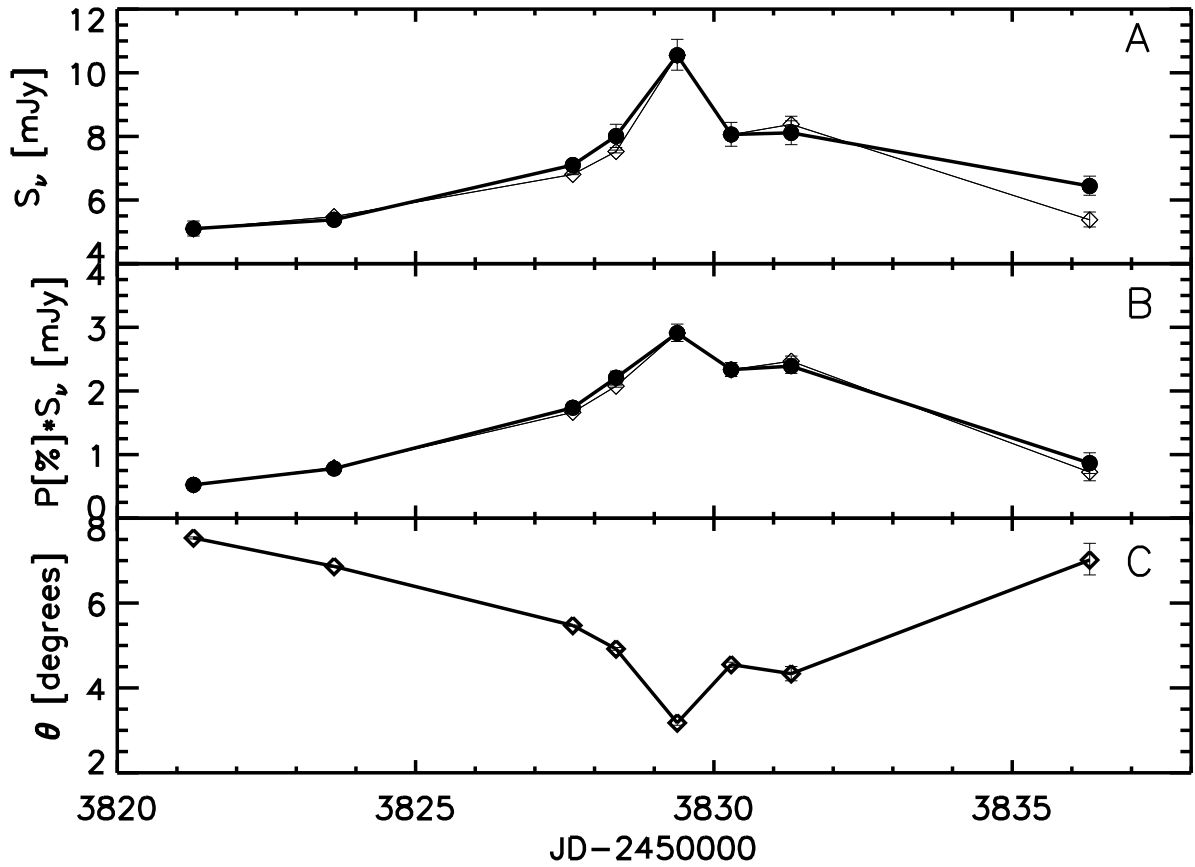}
\caption{Model fit to 2006 local flux density peak.  {\it A:} Observed flux density at R waveband ({\it filled circles}) and modeled flux density ({\it open diamonds}).  {\it B:} Observed polarized flux at R waveband ({\it filled circles}) and modeled polarized flux density ({\it open diamonds}). {\it C:} Modeled viewing angle of the jet in degrees.  Julian date 2453820.5 corresponds to 2006 March 26.
\label{fig9}}
\end{figure}

Figure \ref{fig9} compares the observed and modeled flux density as well as the observed and modeled polarized flux density, and plots the required viewing angle.  We demonstrate in Figure \ref{fig9}, panels A and B, the excellent fit of the spine and sheath model to the observed flux density and polarized flux density.  Following the peak, we observe a decline in flux density combined with a delayed decline in percentage polarization.  We interpret this as a reversal of viewing angle over a period of 7 days from $3.18\pm0.06\degr$ to $7.0\pm0.4\degr$, as shown in Figure \ref{fig9}, panel C.  The optical EVPA (Figure \ref{fig4}) swings from $6.7\degr$ to $-37\degr$ throughout the rise and fall of the flux density.  Based on the previously stated relationship between $\chi$(core) and jet direction, this corresponds to the spine tracing an arc from projected position angle $-83\degr$ to $-127\degr$ over a 2-week period, with the swing in $\chi_{\rm opt}$ providing further evidence for rapid, erratic behavior in jet direction.  This projected arc swung toward, then away from, the line of sight, resulting in the decrease, then increase, of the viewing angle, and thus the increase and decrease observed in polarization and flux density. 

We can check the feasibility of such a spine and sheath structure by examining the distance downstream necessary for such a boundary layer to form in the presence of a velocity mismatch.  For this purpose, we use a formulation from \citet{jor07} for shearing from transverse velocity gradients.  This equation describes the amount of alignment of magnetic field expected by such a gradient, assuming that the field is initially turbulent.  Given the speeds of flow in the spine and sheath, $\beta_{\rm spine}$ and $\beta_{\rm sheath}$, we can calculate the fractional alignment $f_{align}$ at a given distance from the initial area of turbulence:  

\begin{equation}
f_{\rm align} \approx \frac{x \beta_{\rm rel}}{\Gamma R} \frac{\Delta z}{R} \ ,
\end{equation}
where $\Gamma$ is the Lorentz factor of the sheath, $\beta_{\rm rel}=(\beta_{\rm spine} - \beta_{\rm sheath})(1-\beta_{\rm spine}\beta_{\rm sheath})^{-1}$, $R$ is the half-width of the jet, $x$ is the size of the turbulent cells, and $\Delta z$ is the distance between the point where the field is completely turbulent and the location of the feature.  We set an upper limit to $R$ based on component widths close to the 43 GHz core, as determined by model fitting of the VLBA data.  We estimate $x$ by first calculating the number of cells required to reduce the maximum polarization of incoherent synchrotron radiation $P_{max}$ to the observed value of $P$ within a single resolution element, $N = (P_{max}/P)^{2}$, where $N$ is the number of turbulent cells within the beam \citep{bur66}.  Then $x$ is of order the beam width divided by $N^{1/3}$, $\sim 0.049$ pc.  As a check on our value for $x$, we calculate the scale size of turbulent cells from the 2-day quasi-periodicity in $P$(optical) shown in Figure \ref{fig5}, transformed to the plasma frame, which gives a value with the same order of magnitude, $x\sim$0.024 pc.  We find that for the modeled ratio of boundary layer area to turbulent spine area, such alignment can occur within 2.5 pc of the core.

Although the model presented above well describes the observed behavior, we must address the issue of a rapidly changing viewing angle, as indicated in Figure \ref{fig9}.  Previous studies of OJ287 have suggested that the viewing angle of the jet is relatively steady over short time intervals, with major excursions only on timescales of years.  According to the binary black hole model of \citet{val06}, the viewing angle of the jet has a mean value of $2.7\degr$ and varies on the order of $0.5\degr$ over 30 years.  This angular variation is too small to describe the variation of viewing angle in our model.  Additionally, \citet{tat04} suggest that the 8 GHz inner jet demonstrates ballistic precession with a period of 11.6 years.  However, this precession is on a time-scale that is orders of magnitude larger than the 10-day variation we model.  To describe variations in polarized flux and $\chi$ such as we observe, \citet{hug98} consider helical modulation of the velocity vector in OJ287 by a disturbance in the jet.  A helical velocity field would produce the modeled change of viewing angle, which swings first toward, then away from, the line of sight.  However, due to physical constraints, this mechanism requires that rotation of the flow occur on a timescale of greater than a year.  The variations we report here are too rapid to be explained by either a slowly precessing jet or velocity vector rotation.  

\begin{figure}
\epsscale{1.00}
\plotone{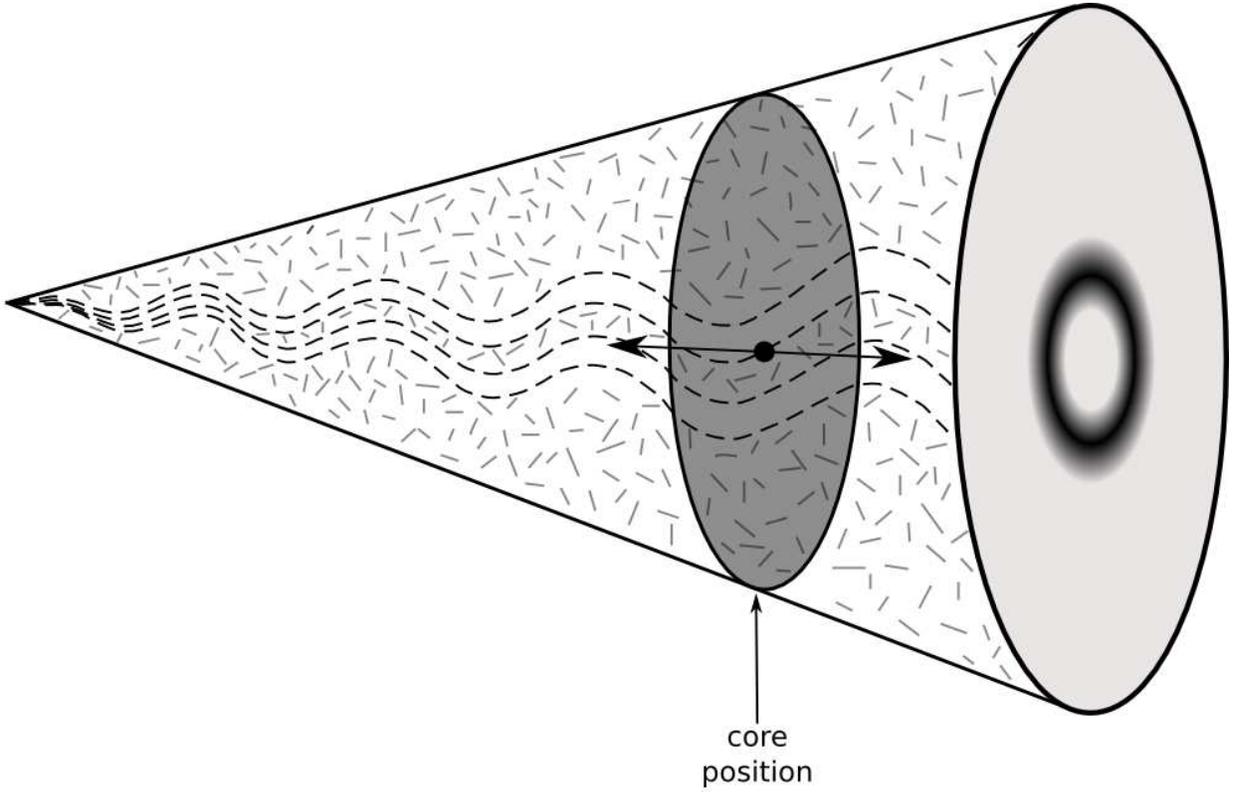}
\caption{Oblique view of jet.  Core position shifts back and forth, causing swings in the velocity vector of plasma in the core.  Line segments represent direction of the magnetic field, which is parallel to the velocity vector in the boundary layer.  Viewing angle of spine and boundary layer vary with time.
\label{drawing}}
\end{figure}

Instead, we propose that the change in viewing angle that we infer is most likely the combination of wiggles caused by an instability and minor fluctuations in the optical depth along the jet.  If the position of the core is defined by the surface at which the optical depth becomes unity \citep{bla79, kon81}, a small change in the optical depth of the jet will move the position of the core upstream or downstream.  As the core, the brightest feature, moves along the jet, it will highlight different parts of the helical velocity field.  Figure \ref{drawing} illustrates the jet structure that we propose.  If the quasi-steady jet is subject to transverse oscillations, perhaps produced by the helical velocity modulation, then the change in core position will cause apparent swings in the direction of the section of the spine that contains the core.  Such movement of the core could be induced by a change in magnetic field and electron density in the wake of the disturbance that creates a new knot, namely component 2 shown in Figures \ref{figa}, \ref{figzoom}, and \ref{fig2}.  This conclusion is supported by \citet{jor02}, in which phase-referencing observations of OJ287 at 43 GHz indicate possible motion of the core during a period of prominent flaring.  {\it We therefore ascribe the swing in viewing angle needed in our model to a temporary change in the position of the core.  This affects the direction of motion of the emitting plasma in the spine without changing the previously-established long-term integrity of the overall jet.}

While we are able to use this model to describe the variable emission in 2006 March, we must also address the emergence of knots from the core.  The polarization vectors of the observed knots tend to align well with the direction of their motion.  For each component, we calculate the difference between its PA and $\chi$.  We find a difference of $3\pm9\degr$ for the combined values of component 2, component 3, and the western component observed only in 2005 October.  This alignment can be explained by the propagation of transverse shock waves \citep{hug85}, which enhance the magnetic field component in the plane of the shock, within the turbulent spine of the inner jet.  The presence of relativistic shocks within the spine of an otherwise longitudinally-aligned spine-sheath system has been previously theorized \citep{gop04} and observed to exist at radio wavelengths \citep[e.g.,][]{att99,gab03a} as well as at optical wavelengths on kiloparsec scales \citep[e.g.,][]{dul07}.  The observed extent of these jets, as well as the consistently high flow speed at large distances, provides evidence for the spine and sheath model.  In addition, optical observations demonstrate the longevity of magnetic structure in the jet, characterized by perpendicular magnetic field in the spine and parallel, aligned magnetic field along the boundary layer. 

At times, when the viewing angle and flux density of the shocked emission are such that the polarized flux density from the shock exceeds that of the sheared boundary layer, the observed $\chi$(core) can switch such that it becomes parallel to the jet.  This parallel $\chi$(core) appeared throughout the 1970's and 1980's with some rapid switching to and from a perpendicular $\chi$ \citep{hag80}.  \citet{sil91} proposed that changes in $\chi$ could be due to temporary intervals of dominance of transversely shocked magnetic field over an underlying, structurally stable longitudinal field in the inner jet.  \citet{rob87} reported a change of $\chi$(core) from $160\degr$ to $99\degr$ in late 1982, coincident with the extrapolated birth of knot K3 \citep{gab89} along PA = $-100\degr$.  Our conclusions are supported by those of \citet{hag91}, in which decomposition of the polarized optical flux of OJ287 revealed two distinct components, one with $\chi=163\degr$ (corresponding to our sheared component) and the other with $\chi=84\degr$ (corresponding to a transverse shock).  In our model, the slow and stationary knots observed intermittently in OJ287, such as component 1 in Figures \ref{figa}, \ref{figzoom}, and \ref{fig2} with $\Gamma = 5.3\pm0.6c$, could be associated with shocks propagating along the slower sheath, as suggested in \citet{kom90}, \citet{lai99}, \citet{chi00}, and \citet{ghi04}.  In addition, component 1 exhibits a difference between PA and $\chi$ of $53\pm15\degr$, suggestive of a more oblique shock than those seen in faster components.  Thus, the spine and sheath dichotomy can produce distinctly separate classes of fast and slow knots as well as the behavior of $\chi$ observed in OJ287 during our campaigns and in previous studies.  These include the fast ($\Gamma$ = 11.6 to 18.0) and slow ($\Gamma$ = 0.4 to 1.3) groups of knots noted in OJ287 by \citet{jor05}.

\subsection{Dependence of Degree of Polarization on Frequency and Flux}
We find a clear discrepancy between the optical and radio core polarization in both the 2005 and 2006 campaigns.  In the 2005 observations, $P$($R$-band) is consistently 2.5 times larger - and $P$($H$-band) 2.1 times larger - than $P$(43 GHz core), while the flux density and degrees of polarization are at high levels.  In 2006, we observe similar low optical, infrared, and radio core percent polarization at a fainter flux density, followed by an increasing ratio between the wavelengths as the flux density and $P$ increase (Figure \ref{fig6}).

We explain this behavior by appealing to a steepening of the spectrum with frequency.  We apply a range of spectral indices to the multiwavelength polarization measurements, and find a good fit for the following wavelengths: $\alpha_{B}=-1.41$, $\alpha_{V}=-1.35$, $\alpha_{R}=-1.29$, $\alpha_{I}=-1.21$, and $\alpha_{H}=-0.81$.  These spectral indices reproduce the observed polarization ratio changes.  The spectral index has little impact on percentage polarization at a high viewing angle, which agrees with the similarity in observed values at low flux density.  We then see a greater impact on polarization at smaller viewing angles, as found in the data, with flatter spectra producing a lower percent polarization.  To check whether the observed steepening in optical index is consistent with synchrotron theory, we examine the indices produced by a uniform synchrotron source for a given index of the electron power-law energy distribution \citep{pac70}.  We find that the steepening can be reproduced with a synchrotron critical frequency of $\sim1.2\times10^{15}$Hz, or $\lambda\sim2500{\rm \AA}$, for an electron energy distribution N(E) $\propto E^{-s}$ with index $s=2.5$.  

\begin{figure}
\epsscale{1.00}
\plotone{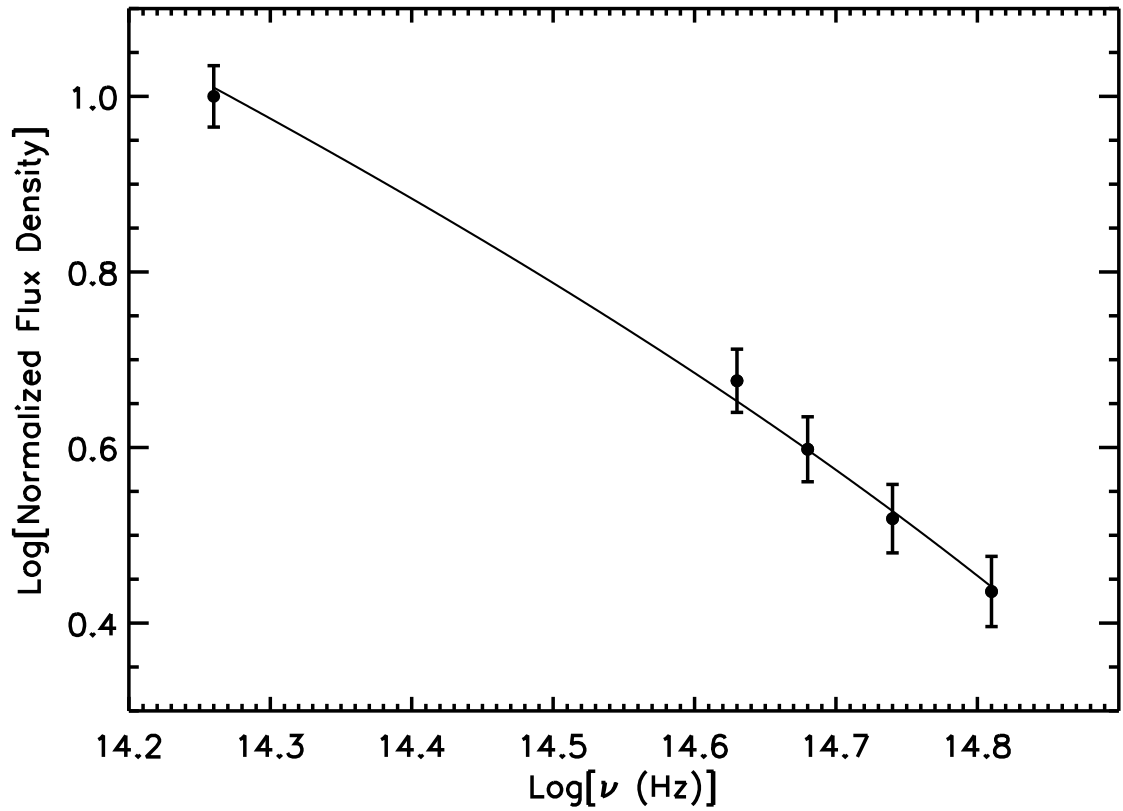}
\caption{Optical-to-infrared spectrum from 2006 campaign.  Displayed above is the average spectrum, normalized to peak at unity in the $H$-band.  The optical spectrum is binned in 4 equal segments of wavelength between 4200${\rm \AA}$ and 7500${\rm \AA}$.  The solid line displays the modeled spectrum.
\label{sedplot}}
\end{figure}

We examine the optical spectra to determine whether this steepening in index was in fact observed.  We are not able to detect such steepening within the optical region given the error in spectral index for individual wavebands, $\sigma$($\alpha$)$\sim\pm0.25$.  In Figure \ref{sedplot}, we display the average optical-to-infrared spectrum overlaid with the model steepened spectrum.  For each epoch at which both infrared and optical measurements are available, we normalize the observed spectrum to log[Flux Density($H$)]=1, then average the spectra to produce the displayed data.  As is shown in Figure \ref{sedplot}, we observe a clear steepening between the infrared to far-red index ($\alpha_{H-I}=-0.9\pm0.1$) and the optical index ($\alpha_{opt}=-1.3\pm0.1$), which is fit well by the model steepened spectrum.  In addition, all five modeled indices agree with the observed values within one standard deviation.  Historically, such steepening between infrared and optical wavelengths has been observed in OJ287 by \citet{rie74}, \citet{wor82}, \citet{hol84}, \citet{sit85}, \citet{bal90}, \citet{gho95}, and \citet{hag98}.

When we extend the spectrum to fit the 43 GHz data, we find that the percentage polarization in the radio requires an optically thin spectral index near zero.  From this, we conclude that the 43 GHz polarization is diluted by emission from an optically thick region.  As described in \citet{gab03}, the agreement between $\chi$(optical) and $\chi$(43 GHz core) suggests that the polarized radio emission is dominated by the optically thin region.  We calculate the necessary dilution by the weakly polarized optically thick zone using the formula derived in \citet{pac70}, under the assumption that the optically-thin spectral index $\sim-0.8$, roughly that inferred for the infrared $H$-band (cf. the spectral energy distribution derived by \citet{imp88}).  From the observed percentage, we determine that $\sim30\%$ of the 43 GHz core emission is optically thick.

\subsection{Polarization Behavior During 2005 Campaign}
Examination of the 2005 campaign results reveals behavior that, while less dynamic, agrees with the above model.  As Figure \ref{fig3} indicates, we observe an upward trend in both optical and 43 GHz core flux, as well as an increase in $P$(optical), accompanied by a swing in $\chi$(optical).  A change in viewing angle of the spine can explain all three observed effects. 

\subsection{Alternative Explanations of the Polarization and Variability}
We consider two alternative explanations for the observed behavior of $\chi$ in the 43 GHz core of OJ287.  In one recent proposal, \citet{gab04} use complex rotation measures and close examination of $\chi$ values to describe the parsec-scale jets of BL Lac objects as containing a helical field in which the toroidal component of magnetic field dominates.  While this description is consistent with the steadiness of $\chi$ in components 2 and 3 as they proceed down the jet, it fails to explain either the dynamical nature of the core or the fast increase in flux and polarization.  In addition, although many BL Lac objects display a $\chi$(core) parallel to the jet direction \citep{gab00, mar02}, our study of OJ287 shows a highly-polarized $\chi$(core) {\it perpendicular} to the jet, a condition that would necessitate extreme, nearly longitudinal pitch angles of the helix.   Historically, the core of OJ287 has high polarization in instances when it maintains a transverse $\chi$ - shown in our data - as well as when it possesses a longitudinal $\chi$ \citep{hag80,hag91}, which is not expected in the helix model.  

Another study of OJ287, by \citet{vic96} using the models of \citet{har87}, suggests that the progression of components along the jet is in fact the motion of shocks following a helical path.  If this path were to change by 90$\degr$ within the resolution of the VLBA, it could very well describe the dichotomy we see between $\chi$ in the core and that of its ejected components, as well as the change in viewing angle as plasma proceeds down the parsec-scale jet.  Nonetheless, we observe no helical behavior in the path of motion of the components within positional errors (see Figure \ref{fig2}), contradicting the model.  Furthermore, \citet{tat04} have examined this model for OJ287 and conclude that it is insufficient to fully describe the data, proposing instead that ballistic precession produces a changing inner jet direction on longer timescales than seen in our observations.  We note, however, that the conclusions of \citet{vic96} described components on scales of several milliarcseconds and on timescales of several years, and are therefore not entirely applicable to the data that we display.

\section{Summary and Conclusions}
The increase of polarization across millimeter to optical wavelengths that our observations reveal in OJ287 leads us to conclude that the optical emission is cospatial with the 43 GHz core.  This conclusion is supported by the similarity in value and behavior with time of the EVPA at different wavebands.  The EVPA of the core has maintained an angle roughly perpendicular to the flow of the jet both historically - except for occasional $90\degr$ flips - and in our observations.  This steadiness implies persistence of a region of velocity shear in the core region.  We have proposed a model in which velocity shear creates an ordered longitudinal component of the magnetic field in the transition region between a fast, turbulent spine of the jet, and a slower, turbulent sheath.  Shock waves emerge along the spine to form superluminal knots, and along the sheath to produce relatively slow, or stationary, features.  We model the core with this spine-sheath structure, with the inner jet containing flux concentrations.  With the existence of velocity vector gradients in the inner jet, changes in the position of the core cause variations in position and viewing angles.  The occasional appearance of shocks, whose emission dominates the flux density, can explain the $90\degr$ jumps in $\chi$(43 GHz core) observed at previous epochs.  

Our observations of OJ287 have revealed a dynamic jet downstream of the core.  It no longer shows relatively slow and steady motions as observed in the 1980's and 1990's, nor does it display non-core stationary features as observed in the late 1990's.  We have uncovered a tendency for the jet to produce knots following different trajectories at intervals of months.  The viewing angle changes that we propose are counter to those called for in studies of slower, systematic viewing angle changes in OJ287.  Rather, they correspond to more rapid fluctuations in the direction of the flow of the emitting plasma within the core.  We speculate that disturbances propagating down the jet cause shifts in the location of the site where the optical depth equals unity, which is identified as the core on VLBA images.  If the direction of the spine is subject to an instability that results in spatial oscillations, the shifting core position would appear as a swing in the direction of the fast spine.  This change in angle between the flow velocity vector and the line of sight would then be responsible for the variability in flux and polarization that we have observed.

We have collected similar data from a number of other blazars during the 2005 October-November campaign and the 2006 March-April campaign.  We are currently analyzing the data to determine the level of generality of the conclusion that polarized optical emission is correlated in variability and polarization properties with the 43 GHz core in blazars.  The result of this analysis, in concert with previously established correlations between optical emission and high-frequency emission, will contribute toward the development of a general model for emission in parsec-scale jets.  This will be described in a summary paper to follow.  

\acknowledgments
Support for this work was provided by the National Science Foundation (NSF) through award GSSP07-0009 from the NRAO.  The effort at Boston University was also supported by NASA Astrophysical Data Analysis Program grant NNX08AJ64G and by the NSF under grant AST 04-06865.  P. S. S. acknowledges support from NASA contract 1256424.  The St. Petersburg team acknowledges support from The Russian Foundation for Basic Research grant 09-02-00092.  The VLBA is an instrument of the NRAO, a facility of the NSF operated under cooperative agreement by Associated Universities, Inc.  This research was conducted in part using the Mimir instrument, jointly developed at Boston University and Lowell Observatory and supported by NASA, NSF, and the W.M. Keck Foundation.

{\it Facilities:} \facility{VLBA}, \facility{SO:Kuiper}, \facility{CrAO}, \facility{Perkins}.

\clearpage
\begin{tabular}{l | l | c | c | c | c | c | c}
\hline
{\bf Epoch} & {\bf Component} & {\bf R (mas)} & {\bf PA ($\degr$)} & {\bf I (Jy)} & {\bf P ($\%$)} & {\bf $\chi\degr$} & {\bf $Chi^{2}_{DOF}$\footnotemark[1]}\\ \hline
\hline
05 Oct. 24 & Core & - & - & 2.01 & 12.7 & 21 & 13.7\\ \hline
 & West Component & 0.08 & -58 & 1.09 & 7.0 & -68 & 5.2\\ \hline
 & Component 1 & 0.32 & -147 & 0.03 & 0.0 & - & 3.1\\ \hline
\hline
05 Oct. 28 & Core & - & - & 2.30 & 11.8 & 16 & 9.6\\ \hline
 & West Component & 0.09 & -84 & 0.77 & 6.1 & -87 & 3.1\\ \hline
 & Component 1 & 0.38 & -154 & 0.02 & 0.0 & - & 2.0\\ \hline
\hline
05 Nov. 2 & Core & - & - & 2.33 & 11.8 & 19 & 9.1\\ \hline
 & West Component & 0.11 & -80 & 0.65 & 7.6 & -83 & 3.9\\ \hline
 & Component 1 & 0.40 & -159 & 0.03 & 0.0 & - & 2.6\\ \hline
\hline
06 Mar. 27 & Core & - & - & 1.55 & 9.7 & -4 & 41.8\\ \hline
 & Component 2 & 0.10 & -90 & 1.35 & 0.0 & - & 28.2\\ \hline
 & Component 1 & 0.54 & -145 & 0.03 & 69.9 & -8 & 20.4\\ \hline
\hline
06 Mar. 31 & Core & - & - & 1.72 & 13.8 & 1 & 15.7\\ \hline
 & Component 2 & 0.12 & -91 & 2.01 & 0.8 & 75 & 4.6\\ \hline
 & Component 1 & 0.60 & -145 & 0.03 & 61.2 & -22 & 4.2\\ \hline
\hline
06 Apr. 5 & Core & - & - & 1.82 & 14.4 & 1 & 16.7\\ \hline
 & Component 2 & 0.10 & -109 & 2.07 & 1.7 & 78 & 4.7\\ \hline
 & Component 1 & 0.56 & -145 & 0.05 & 31.5 & -28 & 4.6\\ \hline
\hline
06 May 16 & Core & - & - & 4.31 & 5.4 & 11 & 36.7\\ \hline
 & Component 2 & 0.21 & -115 & 0.31 & 2.9 & 60 & 7.8\\ \hline
 & Component 1 & 0.52 & -146 & 0.04 & 24.9 & -23 & 7.6\\ \hline
\hline
06 Jun. 17 & Core & - & - & 4.75 & 4.0 & 25 & 32.3\\ \hline
 & Component 2 & 0.27 & -110 & 0.12 & 16.0 & 66 & 9.7\\ \hline
 & Component 1 & 0.53 & -145 & 0.08 & 10.4 & 11 & 8.5\\ \hline
\hline
06 Aug. 10 & Core & - & - & 1.09 & 5.7 & 40 & 25.5\\ \hline
 & Component 3 & 0.10 & -78 & 1.27 & 0.8 & -61 & 13.9\\ \hline
 & Component 2 & 0.41 & -117 & 0.14 & 6.0 & 52 & 9.1\\ \hline 
 & Component 1 & 0.57 & -144 & 0.12 & 10.3 & -24 & 7.1\\ \hline
\hline
06 Oct. 5 & Core & - & - & 2.14 & 1.8 & 12 & 3.4\\ \hline
 & Component 3 & 0.21 & -98 & 0.24 & 4.6 & 80 & 1.9\\ \hline
 & Component 1 & 0.67 & -143 & 0.06 & 29.0 & -32 & 1.0\\ \hline
\end{tabular}

\footnotetext [1]{$Chi^{2}_{\rm DOF}$ = $Chi^{2}$/(2 x no. visibilities - no. modelfitting parameters) of the model containing the respective component plus the components above it.}

\clearpage

\clearpage

\clearpage

\clearpage

\clearpage

\clearpage

\clearpage

\clearpage

\clearpage

\clearpage

\end{document}